\documentclass[pra,twocolumn,amsmath,amssymb,floatfix,reprint,footinbib,superscriptaddress,longbibliography,showkeys]{revtex4-1}
\usepackage{dcolumn}
\usepackage{graphicx}
\usepackage{mathrsfs}
\usepackage{mdwlist}
\usepackage{subfigure}
\usepackage{booktabs}
\usepackage{amsmath}
\usepackage{dsfont}
\usepackage{amstext}
\usepackage{amssymb}
\usepackage{amsbsy}
\usepackage{bbm}
\usepackage{amsthm}
\usepackage{graphicx}
\usepackage{textcomp}
\usepackage{multirow}
\usepackage{color}
\usepackage{xcolor}
\usepackage[colorlinks,citecolor=blue]{hyperref}

\definecolor{Dgreen}{RGB}{0, 100, 0}
\usepackage{url}
\usepackage[colorlinks]{hyperref}
\hypersetup{%
	plainpages=true,
	breaklinks=true,  
	hypertexnames=false,  
	pageanchor=true,
	colorlinks=true,
	linkcolor={blue},
	citecolor={red},
	urlcolor={blue},
	anchorcolor={black}
}

\begin{document}
	
	\title{Enhanced-Fidelity Ultrafast Geometric Quantum Computation \\ Using Strong Classical Drives}
	\author{Ye-Hong Chen}
	\affiliation{Theoretical Quantum Physics Laboratory, Cluster for Pioneering Research, \\ RIKEN, Wako-shi, Saitama 351-0198, Japan}
	\affiliation{Quantum Information Physics Theory Research Team, Center for Quantum Computing, RIKEN, Wako-shi, Saitama 351-0198, Japan}
	\affiliation{Fujian Key Laboratory of Quantum Information and Quantum Optics, \\ Fuzhou University, Fuzhou 350116, China}
	\affiliation{Department of Physics, Fuzhou University, Fuzhou 350116, China}

	\author{Adam Miranowicz}
	\affiliation{Theoretical Quantum Physics Laboratory, Cluster for Pioneering Research, \\ RIKEN, Wako-shi, Saitama 351-0198, Japan}
	\affiliation{Institute of Spintronics and Quantum Information,
		Faculty of Physics, Adam Mickiewicz University, 61-614 Pozna\'n, Poland}
	
	\author{Xi Chen}
	\affiliation{Department of Physical Chemistry, University of the Basque Country UPV/EHU, Apartado 644, 48080 Bilbao, Spain}
	\affiliation{EHU Quantum Center, University of the Basque Country UPV/EHU, Barrio Sarriena, s/n, 48940 Leioa, Spain}
	
	\author{Yan Xia}
	\affiliation{Fujian Key Laboratory of Quantum Information and Quantum Optics, \\ Fuzhou University, Fuzhou 350116, China}
	\affiliation{Department of Physics, Fuzhou University, Fuzhou 350116, China}
	
	\author{Franco Nori}
	\affiliation{Theoretical Quantum Physics Laboratory, Cluster for Pioneering Research, \\ RIKEN, Wako-shi, Saitama 351-0198, Japan}
	\affiliation{Quantum Information Physics Theory Research Team, Center for Quantum Computing, RIKEN, Wako-shi, Saitama 351-0198, Japan}
	\affiliation{Department of Physics, University of Michigan, Ann Arbor, Michigan 48109-1040, USA}

	\date{\today}

	\begin{abstract}
		{We propose a general approach to implement ultrafast nonadiabatic geometric single- and two-qubit gates by employing counter-rotating effects}. 
		This protocol is compatible with most optimal control methods used in previous rotating-wave approximation (RWA) protocols;
		thus, it is as robust as (or even more robust than) the RWA protocols.
		Using counter-rotating effects allows us to apply strong
		drives.
		Therefore, we can improve the gate speed by 5--10 times compared to the RWA 
		counterpart for implementing high-fidelity ($\geq 99.99\%$) gates. Such an ultrafast evolution (nanoseconds, even picoseconds)
		significantly reduces the influence of decoherence (e.g., the qubit dissipation and dephasing).
		Moreover, because the counter-rotating effects no longer induce a gate infidelity (in both the weak and strong driving regimes), we can achieve a
		higher fidelity compared to the RWA protocols.
		Therefore, in the presence of decoherence, one can implement 
		ultrafast geometric quantum gates with $\geq 99\%$ fidelities.
	\end{abstract}
	
	
	\maketitle

\section{Introduction}
Quantum computers promise to drastically outperform classical
computers on certain problems, such as factoring, (approximate)
optimization, boson sampling, or unstructured database searching
 \cite{Aaronson2011,Farhi2022Quantum,HidaryBook,Kockum2019,Kjaergaard2020,ZhongSci2020,LiptonBook}.
To realize a quantum computer, one key ingredient is to realize high-fidelity quantum gates, especially, single- and two-qubit gates.
This is because any unitary transformations, including multiqubit gates, can
be decomposed into a series of single qubit operations along
with universal two-qubit gates (see e.g, \cite{Nakamura1999,NakamuraPRL2001,MakhlinRMP2001,YouPRL2002,YouPRB2003,WeiPRB2005,Buluta2011,Gu2017,Kockum2019,BruzewiczApr2019,HuangSCIS2020,Kjaergaard2020,HuangSCIS2020,CaiFR2021}).
However, gate infidelities, due to decoherence, impede the physical implementation of large-scale
quantum computers \cite{HidaryBook}. Many efforts have been made to solve the above problems. 
Among them, quantum geometric gates  \cite{Zanardi1999,Ekert2000Jmo,Jones2000Nat,Wang2002Prl,ZhuPRL2003,ZhuPRA2005,Filipp2009Prl,SjqvistNJP2012,Berger2013Pra,XuPRL2012,SjqvistIJQC2015,ZhengPRA2016,Zhao2017Pra,XuPRL2018,LiuPRL2019,Zeng2019Pra,XuPRL2020,LiuPRR2020,DongPRXq2021,LiuPRR2021,ZhangarXiv2021}, based on Abelian \cite{Berry1984} and non-Abelian \cite{Aharonov1987,Anandan1988}
geometric phases,
have become promising because geometric phases are
determined by the global properties of the evolution
paths and are intrinsically noise-resilient
against certain types of local noises.
{
For instance, it has been demonstrated \cite{Dechiara2003Prl,Filipp2009Prl,Berger2013Pra} that geometric phases are robust against fluctuations described as Ornstein-Uhlenbeck processes, i.e., stationary, Gaussian, and Markovian noises which have a Lorentzian spectrum.}

However, a geometric gate is relatively slow because it consumes
extra resources to eliminate dynamical phases. Noise 
may accumulate in a slow evolution, thus reducing the gate fidelity.
Though some efforts have been made \cite{SjqvistNJP2012,SongNJP2016,LiuPRL2019,DuAQT2019,LiFP2021,ChenPRR2021,SetiawanPRXQ2021,ShenPRApp2021}, 
only little progress has been achieved in speeding up the gates.
In particular, working under the rotating-wave approximation (RWA), 
it is challenging to accelerate the gates using
finite-interaction strengths, which are much smaller than the qubit transition frequency \cite{ZhangarXiv2021}.

{The above problem motivates us to employ counter-rotating terms (which are usually neglected in many previous protocols) for nonadiabatic geometric quantum computation (NGQC),} so that one can apply strong interactions to achieve ultrafast and high-fidelity computation \cite{WangPRA2016,WangSR2017,WangPRAp2020}. 

In this paper, we propose a general approach for ultrafast NGQC
using driven two-level systems. Using strong drivings effectively
shortens the gate time to nanoseconds (even picoseconds), and, thus, significantly reduces the influence of decoherence \cite{CasasJPA2001,AshhabPRA2007,LuPRA2012,GoldmanPRX2014,YanPRA2015,DengPRA2016,HanPRA2020}.
The effective Hamiltonian obtained by the Floquet theory \cite{ShirleyPR1965,TuorilaPRL2010} possesses a RWA-like form. Thus it is compatible with most optimal control methods \cite{TorosovPRL2011,HerterichPRA2016,ZhaoPRA2017,ZhangPRA2019,DridiPRL2020,LiuPRL2019,LiuPRR2020,LiuPRR2021,DongPRXq2021,SetiawanPRXQ2021}
which have been 
applied under the RWA, such as the recently developed methods of super-robust geometric control \cite{LiuPRR2021} and doubly geometric quantum control \cite{DongPRXq2021}.

The proposed protocol can avoid the negative effects caused by the counter-rotating (CR) interactions, 
including the Bloch-Siegert (BS) shift, which
may shift the qubit transition frequency and induce additional systematic noise in the system.
Thus, this protocol can suppress systematic noise better than the usual RWA counterpart. 
We also generalize the protocol to, e.g., two-qubit holonomic gates, using
strong qubit-qubit couplings. 
{Therefore, this protocol can be a possible replacement for conventional RWA methods, and to improve the speed and fidelity of the NGQC.
Our approach is different from previous non-RWA protocols, e.g., Refs. \cite{Mousolou2014NJP,WangPRA2016,WangSR2017,Mousolou2017PRA,Sorensen2018PRA,WangPRAp2020}, which
work for specific targets.
}

\begin{figure*}
	\centering
	\scalebox{0.6}{\includegraphics{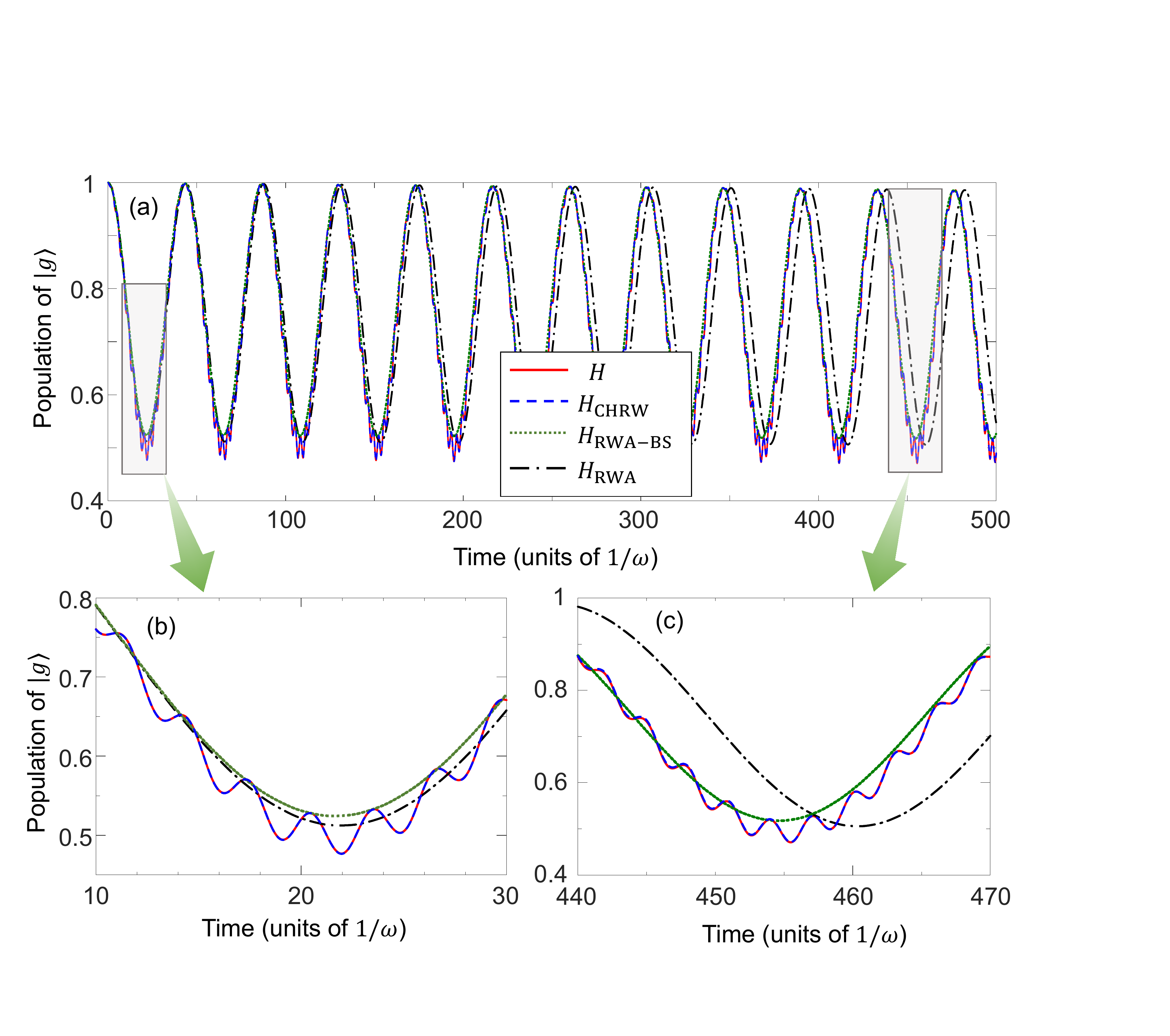}}
	\caption{Population of the ground state $|g\rangle$ in the laboratory frame governed by the Hamiltonians under different approximation protocols: $H(t)$ in Eq.~(\ref{eq1}), $H_{\rm{CHRW}}(t)$ in Eq.~(\ref{eq5}), $H_{\rm{RWA}}(t)$ in Eq.~(\ref{eq2}), and $H_{\rm{RWA-BS}}(t)$ in Eq.~(\ref{eqRWABS}). We choose parameters $\tilde{\Delta}_{q}=\Delta_{q}=0.1\omega$, $\Omega_{0}=0.1\omega$, and $\Omega_{1}=0$. Note that each curve is calculated using the corresponding Hamiltonian after its transformation back to the laboratory frame.
	}
	\label{fig1}
\end{figure*}

{\section{Effective Hamiltonian under strong drives}}
We consider a two-level atom (with ground state $|g\rangle$ and excited state $|e\rangle$) driven
by a two-tone drive with the same frequency $\omega$, different amplitudes $\Omega_{n}(t)$, and phases $\phi_{n}$ ($n=0,1$). 
The Hamiltonian reads (hereafter, $\hbar=1$)
\begin{align}\label{eq1}
  H(t)=\frac{\omega_{q}}{2}\sigma_{z}+\sum_{n=0,1}\Omega_{n}(t)\cos(\omega t+\phi_{n})\sigma_{x},
\end{align}
where $\sigma_{\eta}$ ($\eta=x,~y,~z,~+,~-$) are Pauli matrices. {For weak drivings (i.e., $\Omega_{n}\ll \omega,\omega_{q}$), 
we can perform 
\begin{align}
	H_{I}=&\exp\left(\frac{i}{2}\omega t\sigma_{z}\right)\left[H(t)-\frac{\omega}{2}\sigma_{z}\right]\exp\left(-\frac{i}{2}\omega t\sigma_{z}\right)\cr
	=&\sum_{n}\left[\frac{e^{i\phi_{n}}\Omega_{n}(t)}{2}\left(1+e^{2i\omega t}\right)\sigma_{-}+{\rm{h.c.}}\right]\cr
	&+\frac{\Delta_{q}(t)}{2}\sigma_{z},
\end{align}
where $\Delta_{q}=\omega_{q}-\omega$ is the detuning. 
The fast-oscillating term with $\exp(\pm2i\omega t)$ can be neglected under the RWA,} and the effective Hamiltonian becomes 
\begin{align}\label{eq2}
  H_{\rm{RWA}}(t)=\frac{\Delta_{q}(t)}{2}\sigma_{z}+\sum_{n}\frac{\Omega_{n}(t)}{2}\left(e^{i\phi_{n}}\sigma_{-}+{\rm{h.c.}}\right).
\end{align}
This Hamiltonian has been widely applied to holonomic computation \cite{ZhangarXiv2021}.
{However, the condition $\Omega_{n}\ll\omega,\omega_{q}$ limits the gate speed. When we choose a relatively strong driving amplitude, e.g., $\Omega_{n}\sim 0.1\omega$,
the neglected CR effect (which includes the BS shift) can induce an infidelity (see the black-dashed-dotted curve in Fig.~\ref{fig1}).} Here, the BS shift calculated by the second-order process is
\begin{align}\label{eqRWABS}
  H_{\rm{BS}}=\sigma_{z}\sum_{n}\frac{\Omega_{n}^{2}(t)}{8\omega}.
\end{align} Accordingly,
the effective Hamiltonian becomes
\begin{align}
  H_{\rm{RWA-BS}}(t)=H_{\rm{RWA}}(t)+H_{\rm{BS}}.
\end{align} 
For simplicity, the RWA-based protocol considering the BS shift is denoted hereafter as the ``RWA-BS'' protocol.
{When this BS shift is considered, the phase mismatch can be fixed (see the green-dotted curve in Fig.~\ref{fig1}). However,
the actual dynamics (red-solid curve)
is still not in good agreement with the effective dynamics (green-dotted curve).}

To implement NGQC {with the CR terms},
we transform the Hamiltonian $H(t)$ with a time-dependent generator \cite{ShirleyPR1965,TuorilaPRL2010}
\begin{align}
S(t)=&\exp\left[i\frac{Z}{2}\sin(\tau)\sigma_{x}\right],\cr
\tau=&\omega t+\phi_{0},
\end{align}
resulting in
{
\begin{align*}\label{eq3}
	H'(t)=&{S(t)}H(t){S^{\dag}(t)}-i{S(t)}\dot{S}^{\dag}(t) \cr
	=&\frac{\omega_{q}}{2}\cos\left[Z\sin(\tau)\right]\sigma_{z}\cr
	&+\frac{\omega_{q}}{2}\sin\left[Z\sin(\tau)\right]\sigma_{y}\cr
	&+\left[\Omega_{0}(t)-\frac{Z}{2}(\omega+\dot{\phi}_{0})\right]\cos(\tau)\sigma_{x}\cr
	&-\frac{\dot{Z}}{2}\sin(\tau)\sigma_{x}
	+\Omega_{1}(t)\cos(\omega t+\phi_{1})\sigma_{x},
\end{align*}
with a real time-dependent parameter $Z$ to be determined below.
The last-line expression in $H'(t)$ can be removed by choosing $\Omega_{1}=\dot{Z}/2$ and $\phi_{1}=\phi_{0}-\pi/2$.
Then, using the identity 
\begin{align}
	\exp\left[iZ\sin(\tau)\right]=\sum_{m=-\infty}^{+\infty}J_{m}(Z)\exp(i m\tau), 
\end{align}
the Hamiltonian $H'(t)$ becomes
\begin{align}\label{eq5a}
	H'(t)=&H'_{0}(t)+H'_{1}(t)+H'_{2}(t), \cr
	H'_{0}(t)=&\frac{\omega_{q}}{2}J_{0}(Z)\sigma_{z},\cr
	H'_{1}(t)=&\tilde{\Omega}_{0}(t)\cos(\tau)\sigma_{x}+\omega_{q}J_{1}(Z)\sin(\tau)\sigma_{y},\cr
	H'_{2}(t)=&\omega_{q}\sum_{m=1}^{\infty}J_{2m+1}(Z)\sin[(2m+1)\tau]\sigma_{y}\cr
	          &+\omega_{q}\sum_{m=1}^{\infty}J_{2m}(Z)\cos(2m\tau)\sigma_{z},
\end{align}
where $J_{m}(Z)$ is the $m$th order
Bessel function of the first kind and
\begin{align}
  \tilde{\Omega}_{0}(t)=\left[\Omega_{0}(t)-\frac{Z}{2}(\omega+\dot{\phi}_{0})\right],
\end{align} 
is the effective driving amplitude.
The Hamiltonian $H'_{2}(t)$ includes all higher-order harmonic terms, which can be neglected for $Z\in[0,1]$ \cite{YanPRA2015}. 
Note that this transformation is also valid for multi-level systems by
defining a suitable generator $S(t)$ \cite{CasasJPA2001,GoldmanPRX2014,HanPRA2020}.
By assuming 
\begin{align}\label{eq6a}
	\tilde{\Omega}_{0}(t)=\omega_{q}J_{1}(Z),
\end{align}
the effective Hamiltonian for the system now reads
\begin{align}
	H_{\rm{eff}}(t)&=H'_{0}(t)+H'_{1}(t)\cr
	&=\frac{\omega_{q}}{2}J_{0}(Z)\sigma_{z}+\tilde{\Omega}_{0}(t)\left(e^{i\tau}\sigma_{-}+{\rm{h.c.}}\right),
\end{align}
which possesses a RWA-like form. Because this Hamiltonian contains some counter-rotating terms which are neglected in the standard RWA protocols,
we denote it as a counter-rotating hybridized rotating wave (CHRW)
Hamiltonian \cite{LuPRA2012,YanPRA2015}. By expanding $\exp[i\omega t\sigma_{z}/2]$, we obtain
\begin{align}\label{eq5}
	H_{\rm{CHRW}}(t)=&e^{i\omega t\sigma_{z}/2}H_{\rm{eff}}(t)e^{-i\omega t\sigma_{z}/2}\cr
	                =&\frac{{\tilde{\Delta}_{q}}(t)}{2}\sigma_z+\tilde{\Omega}_{0}(t)\left[e^{i\phi_{0}}\sigma_{-}+{\rm{h.c.}}\right],
\end{align}
which takes the same form as Eq.~(\ref{eq2}) assuming $n=0$. Here, 
\begin{align}
	\tilde{\Delta}_{q}(t)=\omega_{q}J_{0}(Z)-\omega,
\end{align} 
is the effective detuning, $\omega_{q}J_{0}(Z)$ is the renormalized transition frequency of the qubit, and
$\tilde{\Omega}_{0}(t)$ is the effective driving amplitude.
The renormalized quantities in the transformed Hamiltonian
are directly induced by the CR effects.
As shown in Fig.~\ref{fig1}, the dynamics of the CHRW Hamiltonian $H_{\rm{CHRW}}(t)$ (after its transformation back to the laboratory frame) is mostly  
the same as that of the actual Hamiltonian $H(t)$.
}

{According to Eqs.~(\ref{eq5a}) and (\ref{eq6a}), the limitation on the
effective driving strength $\tilde{\Omega}_{0}(t)$ is
\begin{align}\label{eq13}
	&\tilde{\Omega}_{0}(t)\ll{\rm Min}\left[\frac{4J_{1}(Z)m\omega}{J_{2m}(Z)}\right]\ll \frac{4\omega J_{1}(1)}{J_{2}(1)}\approx 15\omega,
\end{align}
In contrast to Eq.~(\ref{eq13}), the limitation on the RWA protocol is
\begin{align}
	\frac{\Omega_{0}(t)}{2}\ll 2\omega.
\end{align}
That is, the CHRW protocol can be $\sim 7.5$ times \textit{faster} than the RWA protocol because the speed of the protocol is inversely proportional to the effective driving strength (i.e., the left-hand sides of the inequalities).
}

\begin{figure}
	\centering
	\scalebox{0.8}{\includegraphics{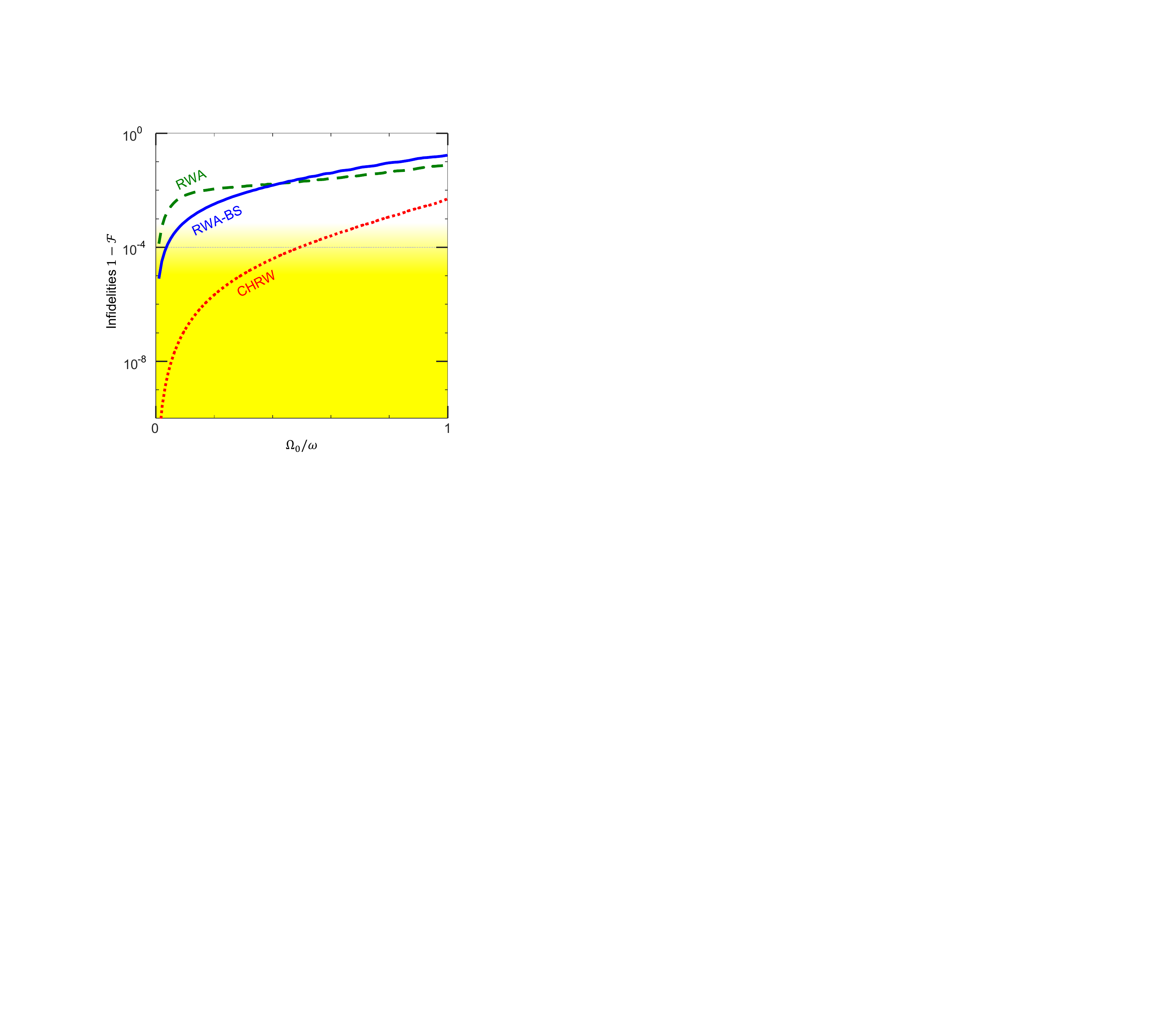}}
	\caption{Infidelities $(1-\mathcal{F})$ averaged over time of the CHRW, 
		RWA-BS, and RWA protocols for $T=100/\Omega_{0}$ with Hamiltonians $H_{\rm{CHRW}}(t)$ in Eq.~(\ref{eq5}), $H_{\rm{RWA}}(t)$ in Eq.~(\ref{eq2}), and $H_{\rm{RWA-BS}}(t)$ in Eq.~(\ref{eqRWABS}), respectively. We choose $\Omega_{0}(t)={\rm{const}}$, $\Delta_{q}(t)=\tilde{\Delta}_{q}(t)=0.1\omega$, and $\phi_{0}=0$. 
		For the RWA and RWA-BS protocols, we choose $\Omega_{1}(t)=0$. 
		Hereafter, all the yellow-shaded areas in the figures correspond to high fidelities ($\gtrsim 99.99\%$); and numerical results are calculated using the Hamiltonian $H(t)$ in Eq.~(\ref{eq1}).
	}
	\label{fig1a}
\end{figure}

To check the range of validity of the above approximations, we define an initial-state-independent fidelity \cite{ZanardiPRA2004,PedersenPLA2007}
$$\bar{F}=\left[\mathrm{Tr}(MM^\dag)
+|\mathrm{Tr}(M)|^2\right]/(D^2+D),$$
where, 
\begin{align}
  M=\mathcal{P}_{{c}}U^\dag_{\rm{eff}}(t)U_{\rm{act}}(t)\mathcal{P}_{{c}},
\end{align} 
and $\mathcal{P}_{{c}}$ ($D=2$) is the projector
(dimension) of the qubit subspace.
The evolution operators $U_{\rm{eff}}(t)$ and $U_{\rm{act}}(t)$ describe the effective and actual dynamical evolutions governed by the approximate Hamiltonian [$H_{\rm{RWA}}(t)$ or $H_{\rm{CHRW}}(t)$] and the total Hamiltonian $H(t)$, respectively.

{Note that the CR effect always influences the system dynamics. To show clearly such influences
in a long-time evolution,} we define an average fidelity,
$\mathcal{F}=\frac{1}{T}\int_{0}^{T}dt \bar{F}$, which averages the fidelities $\bar{F}$ over
time, where $T$ is the total evolution time. 
{This average fidelity evaluates well the error caused by the CR effect.
Moreover, because $U_{\rm{act}}(t)$ describes a set of universal quantum gates,
$\bar{F}$ is also the average fidelity of this set of gates.}

For {$\mathcal{F}= 1$}, the effective dynamics is exactly
the same as the actual one.
Using this definition, in Fig.~\ref{fig1a}, we show that for $\Omega_{0}(t)\sim \omega/2$ (a strong driving), the CHRW protocol (see the red-dotted curve) is valid 
to describe the system dynamics, while the RWA (see the green-dashed curve) is invalid even when the BS shift is considered (blue-solid curve).
Such a strong driving can significantly accelerate the evolution, allowing
ultrafast quantum computation. 
Note that the BS shift obtained by the second-order process is valid only for $\Omega_{0}(t)\ll \omega$. 
For $\Omega_{0}(t)>\omega/2$, it may induce a greater infidelity [see the blue-solid curve in Fig.~\ref{fig1}(a)] 
even compared to the RWA protocol. 

\vspace{5mm}

\section{Implementing fast nonadiabatic geometric gates}
Obviously, $H_{\rm{CHRW}}(t)$ in Eq.~(\ref{eq5})
has exactly the same form as $H_{\rm{RWA}}(t)$ in Eq.~(\ref{eq2}). 
Therefore, the proposed protocol is compatible with the majority of the pulse-design methods \cite{Wang2002Prl,ZhuPRL2003,Zhao2017Pra,Zeng2019Pra,LiuPRR2020,DongPRXq2021,LiuPRR2021}, which have been 
applied for geometric quantum computation under the RWA.
{For a cyclic evolution, we can choose the gate time $T=k\pi/\omega$ ($k=1,2,3,\ldots$) and $S(0)=S(T)=1$, so that
the unitary transformations do not affect the geometric property of the evolution \cite{Berry1984,Aharonov1987,Anandan1988}.}

{
	\renewcommand\arraystretch{1.2}
	\begin{table}
		\centering
		\caption{Parameters used for the examples implementing single-qubit gates.}
		\label{tab1}
		\begin{tabular}{p{1.5cm}<{\centering}p{1.5cm}<{\centering}p{1.5cm}<{\centering}p{1.5cm}<{\centering}p{1.5cm}<{\centering}}
			\hline
			\hline
			Gate & $\alpha(0)$ & $\beta(0)$ & $\Theta_{+}(T)$ & $\Lambda$  \\
			\hline
			NOT & $\pi/2$ & $\pi/2$ & $\pi/2$  & $0.8089$ \\
			Hadamard & $\pi/2$ & $\pi/4$ & $\pi/2$  &  $0.3867$ \\
			Phase-$\pi$ & $0$ & $0$ & $\pi/2$  & $1.4669$ \\
			\hline
			\hline
		\end{tabular}
	\end{table}
}

\begin{figure}[b]
	\centering
	\scalebox{0.35}{\includegraphics{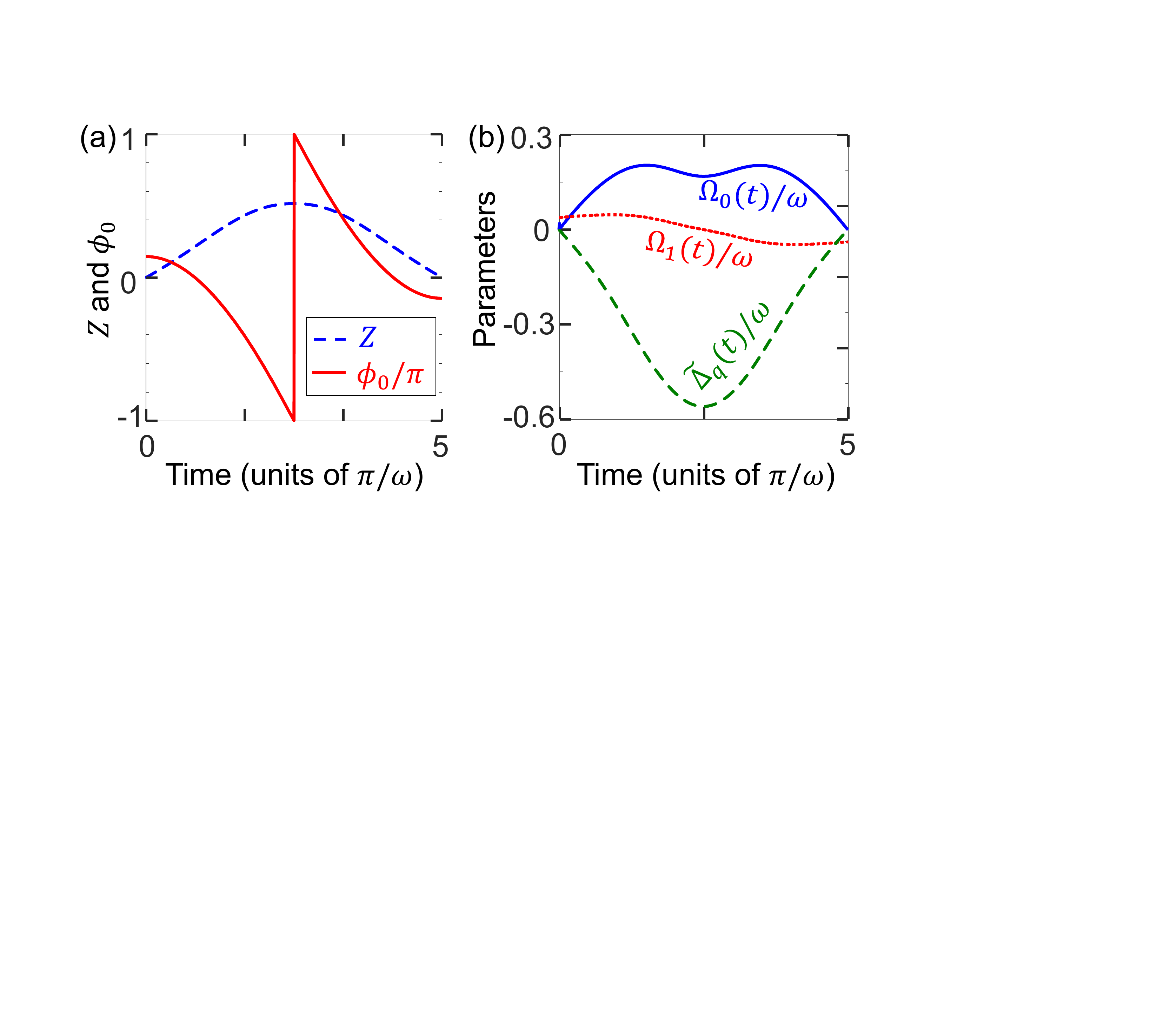}}
	\caption{Implementations of the Hadamard gate using the parameters in Table \ref{tab1}. 
		The parameters shown in (a) and (b) are for the CHRW protocol when $T=5\pi/\omega$. The driving amplitude $\Omega_{0}(t)$
		is comparable to the qubit transition frequency $\omega_{q}$ with these parameters.
	}
	\label{fig1bc}
\end{figure}

According to the Lewis-Riesenfeld theory \cite{LewisJMP1969}, the evolution of the system governed by $H_{\rm{CHRW}}(t)$ can be
described as \cite{ChenPRA2011,ChenPRL2012,IbanezPRL2012,TorronteguiAamop2013,GueryRMP2019} (see more details in Appendix \ref{A1}),
\begin{align}\label{eq11}
 |\phi_{+}(t)\rangle=e^{i\mathcal{R}_{+}(t)}\left[ie^{-i\alpha}\sin(\beta/2),~\cos{(\beta/2)}\right]^{\rm{T}},
\end{align}
or its orthogonal counterpart  
\begin{align}\label{eq12}
|\phi_{-}(t)\rangle=e^{i\mathcal{R}_{-}(t)}\left[\cos(\beta/2),~ie^{i\alpha}\sin{(\beta/2)}\right]^{\rm{T}}.
\end{align} 
Here, $\mathcal{R}_{\pm}(t)$ are the Lewis-Riesenfeld phases, including dynamical and geometric phases, $\alpha$ and $\beta$ are auxiliary parameters to be determined below, and the 
superscript ``T'' is the transposition operator. 

\begin{figure}
	\centering
	\scalebox{0.38}{\includegraphics{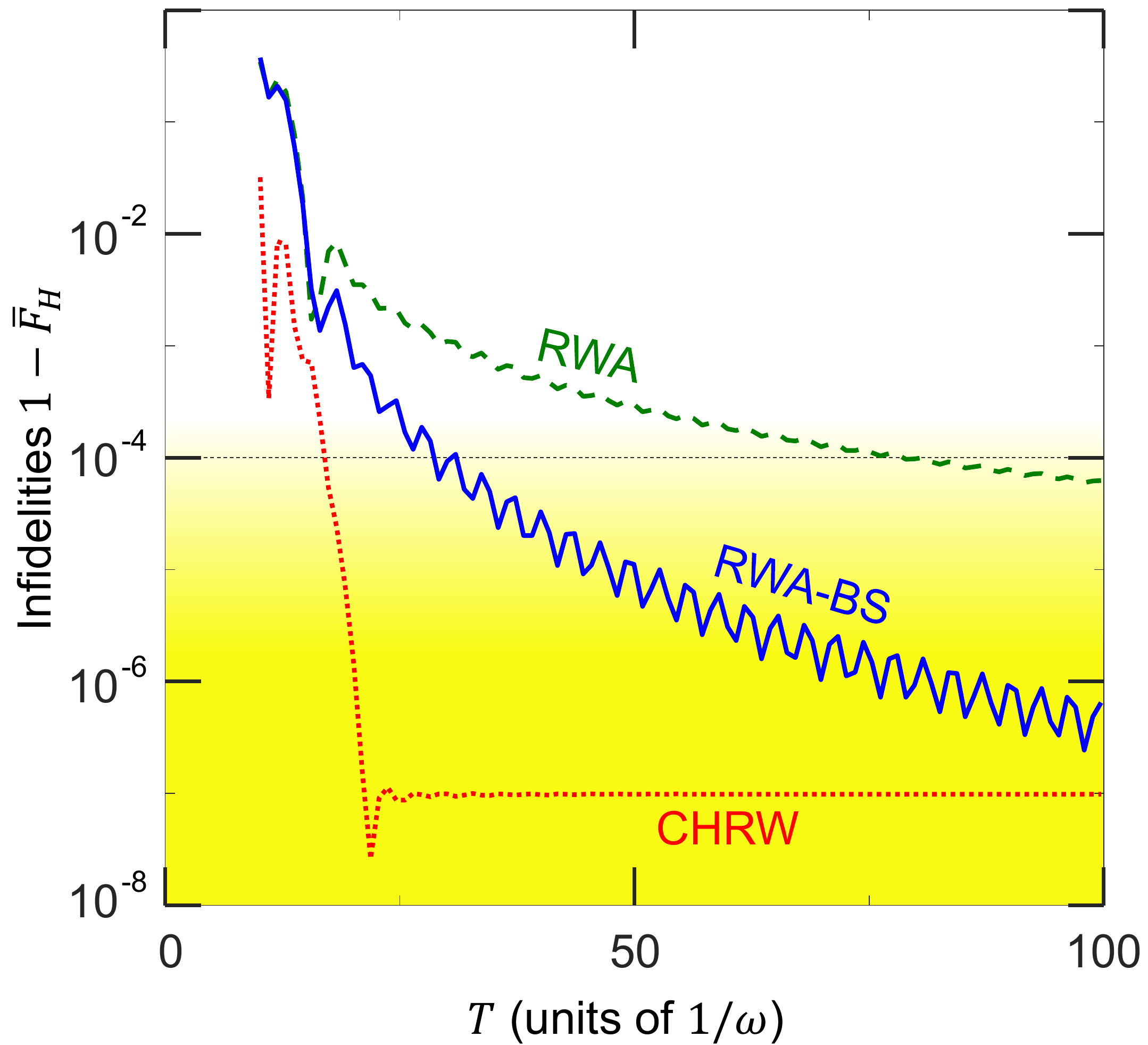}}
	\caption{ Comparisons of infidelities ($1-\bar{F}_{H}$) for the three protocols. For a fixed gate time $T$, the RWA protocol has the highest infidelities, while the CHRW protocol has by far the lowest infidelities.
		For the RWA and RWA-BS protocols, we choose $\Omega_{1}(t)=0$. 
	}
	\label{fig1d}
\end{figure}

To eliminate the dynamical phase, we  
can choose the parameters: $\tilde{\Delta}_{q}(t)=-\dot{\alpha}\sin^{2}\beta$, and
\begin{align}\label{eq6}
  \tilde{\Omega}_{0}(t)\cos{\phi_{0}}=&\frac{1}{4}\left[\dot{\alpha}\sin(2\beta)\sin\alpha-2\dot\beta\cos\alpha\right], \cr
  \tilde{\Omega}_{0}(t)\sin{\phi_0}=&\frac{1}{4}\left[\dot{\alpha}\sin(2\beta)\cos\alpha+2\dot\beta\sin\alpha\right],
\end{align}
resulting in 
\begin{align}
  \langle\phi_{\pm}(t)|H_{\rm{CHRW}}(t)|\phi_{\pm}(t)\rangle=0.
\end{align}
Moreover, the equations of
motion for the geometric phases read 
\begin{align}
  \dot{\Theta}_{\pm}(t) =\pm\dot{\alpha}\sin^{2}\left({\beta}/{2}\right).
\end{align}
Hence, after a cyclic evolution, which is obtained by choosing $\alpha(T)=\alpha(0)\pm 2n_{\alpha}\pi$ and $\beta(T)=\beta(0)\pm 4n_{\beta}\pi$ [$n_{\alpha(\beta)}=0,1,2,\ldots$], the evolution operator at the gate time $T$ becomes
\begin{align}
U_{\rm{eff}}(T)=\sum_{k=\pm}\exp{[i\Theta_{k}(T)]}|\phi_{k}(0)\rangle\langle\phi_{k}(0)|,
\end{align}
which is a universal single-qubit gate.

In this paper, we focus on how the CR effects can shorten the gate time.
Therefore, for simplicity, we choose time-dependent parameters
\begin{align}\label{eq8}
\alpha=&\alpha(0)+\pi\left[1-\cos\left({\pi t}/{T}\right)\right],\cr
\beta=&\beta(0)+\Lambda \sin^{2}\left({\pi t}/{T}\right),
\end{align}
{so that $S(0)=S(T)=1$,
where the parameter $\Lambda$ is numerically obtained according to the geometric phases
$\Theta_{\pm}(T)$.}
The parameters chosen for implementing some single-qubit gates are listed in Table~\ref{tab1}.
Thus, substituting Eq.~(\ref{eq8}) into Eq.~(\ref{eq6}), we can numerically obtain 
the expressions for $Z$ and $\phi_{0}$, and, afterwards, the driving amplitudes $\Omega_{0}(t)$ and $\Omega_{1}(t)$.

{According to the fiber bundle theory, different frames can have all well-defined geometric quantities \cite{Berry1984,Aharonov1987,Anandan1988}. Indeed, the geometric phases defined in different frames satisfy the same property, which is essential to the definition of geometric phases, i.e., they are  invariant under distinct connections and gauge potentials.}

For instance, to implement the Hadamard gate using the parameters listed in Table \ref{tab1}, 
the numerical solutions for $Z$ and $\phi_{0}$ are shown in Fig.~\ref{fig1bc}(a).
Accordingly, we show the driving amplitudes and detuning 
in Fig.~\ref{fig1bc}(b). For $T=5\pi/\omega$, the peak value of the driving amplitude
$\Omega_{0}(t)$ is $\sim0.15\omega$. With such a strong driving, the RWA becomes invalid
to obtain a high-fidelity ($\bar{F}_{H}\gtrsim 99.99\%$) Hadamard gate [see the green-dashed and blue-solid curves in Fig.~\ref{fig1d}]. 
Note that a functional quantum gate should be very precise, typically with a
relative error $\lesssim10^{-4}$ [see the yellow-shaded area 
in Fig.~\ref{fig1d}] \cite{SteanePRSLA1996}. 

In contrast to this, the CHRW protocol can implement
high-fidelity quantum gates using strong drivings. As a result, 
the gate time of the Hadamard gate can be shortened to $T\sim5\pi/\omega$. 
Considering an implementation of the CHRW protocol using natural or artificial atoms,
the driving frequency is $\omega\sim2\pi\times 5~{\rm{GHz}}$ and the gate time is only
$T\sim 0.5~{\rm{ns}}$.

Note from Eqs.~(\ref{eq6}) and (\ref{eq8}) that 
the effective driving amplitude $\tilde{\Omega}_{0}(t)$ is inversely proportional to the gate time $T$.
Therefore, instead of discussing the driving amplitude, analyzing the gate time 
can highlight the advantages (e.g., speed) of the CHRW protocol.
For implementing various quantum gates, the comparisons of the gate speeds for the CHRW, RWA-BS, and RWA protocols are shown in Table~\ref{tab2}.
Generally, the shortest gate times to achieve high-fidelity gates for the CHRW protocol are 
5--10 times shorter than those for the RWA protocol. The RWA-BS protocol also can improve the 
gate speed by 3--4 times, compared to the RWA protocol. 
These indicate that using the CR effects can effectively improve the gate speed for holonomic computation.
For simplicity, the following discussions focus on the Hadamard gate.

{
	\renewcommand\arraystretch{1.2}
	\begin{table}
		\centering
		\caption{The shortest gate time $T$ (in units of $\pi/\omega$) required to implement some high-fidelity  ($\bar{F}\geq 99.99\%$) gates for different protocols.}
		\label{tab2}
		\begin{tabular}{p{1.5cm}<{\centering}p{1.5cm}<{\centering}p{1.5cm}<{\centering}p{1.5cm}<{\centering}p{1.6cm}<{\centering}}
			\hline
			\hline
			Protocol & NOT & Hadamard & Phase-$\pi$ & CNOT-like\\
			\hline
			CHRW & $\sim 5$ & $\sim 6$ & $\sim 6$ & $\sim 5$ \\
			RWA-BS & $\sim 8$ & $\sim 10$ & $\sim 10$ & $\sim 8$\\
			RWA & $\sim 34$ & $\sim 27$ & $\sim 48$ & $\sim 25$  \\
			\hline
			\hline
		\end{tabular}
	\end{table}
}

\section{Robustness against parameter imperfections}
\begin{figure}
	\centering
	\scalebox{0.45}{\includegraphics{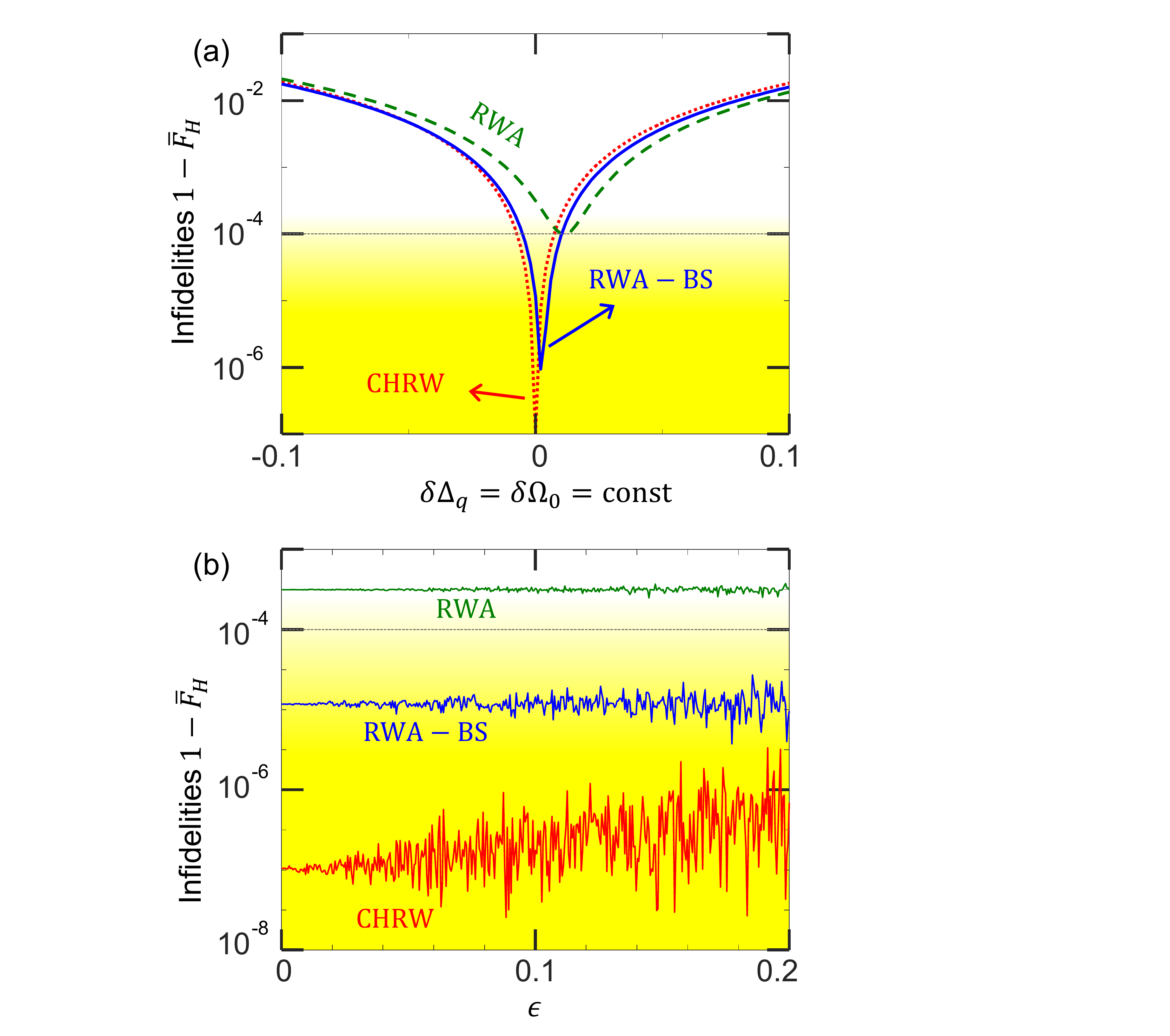}}
	\caption{Gate infidelities ($1-\bar{F}_{H}$) of the CHRW, RWA-BS, and RWA protocols for the Hadamard gate. (a) In the presence of dynamical noise [i.e., $\delta\Delta_{q}(t)=\delta\Omega_{0}(t)$ are constants]. (b) In the presence of stochastic noise [i.e., $|\delta\Delta_{q}(t)|,|\delta\Omega_{0}(t)|\leq \epsilon$ are random numbers]. We choose the same gate time $T=16\pi/\omega$ for the three protocols, so that the driving amplitudes for different protocols are similar to each others. Other parameters are listed in Table~\ref{tab1}.
	}
	\label{fig2}
\end{figure}

Imperfections in the drives are a major source of noise for the discussed system.
The parameter {$X\in\left[\Omega_{0,(1)}(t),\Delta_{q}(t),\phi_{0,(1)},T\right]$} with these imperfections should
be corrected as $X'=X\left(1\pm{\delta X}\right)$, where $\delta X$ denotes the noise rates.
For systematic noise, $\delta X$ is a constant. 
For simplicity, we consider that the noise rates for different parameters are the same,
i.e.,  $\delta\Delta_{q}(t)=\delta\Omega_{0}(t)$.
In the presence of systematic noise, the gate infidelities
for the three protocols are shown in Fig.~\ref{fig2}(a), when choosing the same gate time $T=16\pi/\omega$.
As shown, for small noise rates, the CHRW protocol can suppress systematic noise much better than the RWA protocol; and 
better than the RWA-BS protocol. 
Therefore, for small noise rates (e.g., $\delta \Omega_{0}(t)=\delta{\delta}_{q}(t)\lesssim\pm0.01$), 
it is still possible to implement quantum gates with fidelities $\gtrsim 99.99\%$. 

For stochastic noise, $\delta X$ becomes a random number. We can assume
$\delta X\in\left[-\epsilon,\epsilon\right]$ and numerically study its influence,
where $\epsilon$ denotes the peak noise rate.
Same as above, we now consider noise in ${\Delta}_{q}(t)$ and $\Omega_{0}(t)$, and show
the gate infidelites $(1-\bar{F}_{H})$ in Fig.~\ref{fig2}(b).
We find that stochastic noise affects the protocols very weakly. 
Such noise decreases the gate fidelities by $\sim10^{-4}$, $\sim 10^{-5}$, and $\sim 10^{-6}$
for the RWA (green curve), 
RWA-BS (blue curve), and CHRW (red curve) protocols, respectively. 
This shows that all the three protocols are robust against stochastic noise; and the CHRW protocol can be more robust than the other two.

\section{Decoherence}
Note that an ultrafast evolution can significantly reduce the 
decoherence of a qubit. 
In the presence of decoherence \cite{Breuerbook2007,Lidarnotes2019}, the system dynamics is described by the master equation
\begin{align}
\dot{\rho}=-i[H(t),\rho]+\gamma \mathcal{D}[\sigma_{-}]\rho+\gamma_{\phi} \mathcal{D}[\sigma_{z}]\rho, 
\end{align}
where
\begin{align}
   \mathcal{D}[o]\rho=o\rho o^{\dag}-\frac{1}{2}\left(o^{\dag}o\rho+\rho o^{\dag}o\right)
\end{align}
is the standard Lindblad superoperator and $\gamma$ ($\gamma_{\phi}$) is the spontaneous emission (dephasing) 
rate. 
The fidelity of an output state $|\phi_{\rm{out}}\rangle$ is defined as
$F_{\rm{out}}=\langle\psi_{\rm{out}}|\rho(T)|\psi_{\rm{out}}\rangle$.
Thus, we redefine the gate fidelity $\bar{F}$ as the average value of $F_{\rm{out}}$
over many possible input states.

\begin{figure}
	\centering
	\scalebox{0.42}{\includegraphics{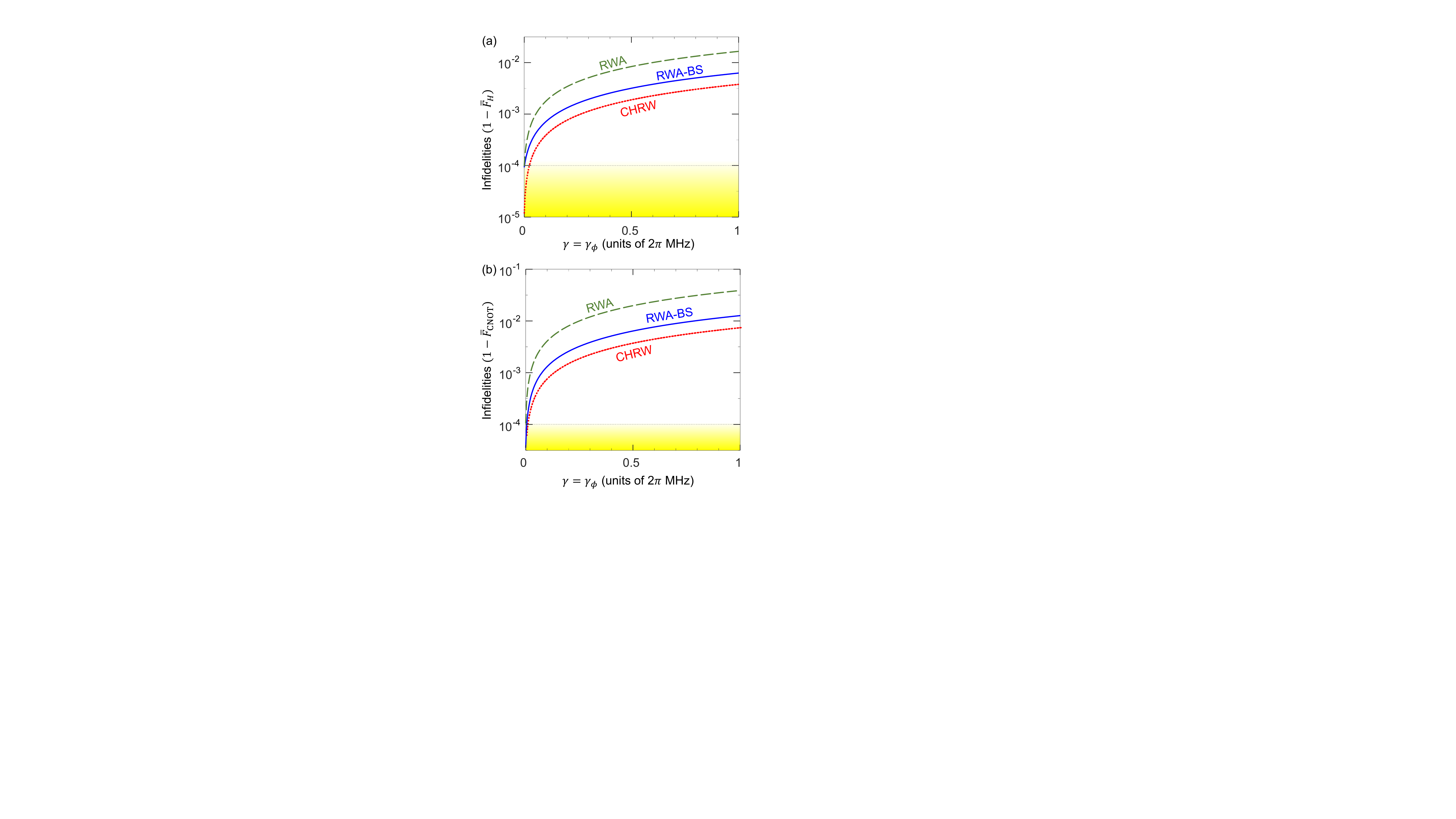}}
	\caption{Gate infidelities: (a) $1-\bar{F}_{H}$ (Hadamard gate) and (b) $1-\bar{F}_{\rm{CNOT}}$ (CNOT-like gate), averaged over 10,000 input states) of the CHRW, RWA-BS, and RWA protocols in the presence of decoherence. We choose the frequency $\omega=2\pi\times 5$ GHz. Other parameters are listed in Tables~\ref{tab1} and \ref{tab2}.  
	}
	\label{figdiss}
\end{figure}

{We can choose the driving frequency $\omega=5$~GHz.
In the presence of decoherence, comparisons of gate fidelities (averaged over 10,000 input states) among the three protocols
are shown in Fig.~\ref{figdiss}(a). 
These input states are uniformly distributed over the Bloch sphere,
which are defined as
\begin{align}
	|\psi\rangle_{\rm in}=\cos\theta_{\rm in}|g\rangle+\sin\theta_{\rm in}e^{i\varphi_{\rm in}}|e\rangle,
\end{align}
where $\theta_{\rm in}\in[0,2\pi]$ and $\varphi_{\rm in}\in[0,2\pi]$ are two parameters determining the input state.
For simplicity, we choose $\theta_{\rm in}$ and $\varphi_{\rm in}$ as arithmetic progressions in the range $[0,2\pi]$ and obtain the 10,000 input states.
}The comparison indicates that the CHRW approach
can achieve much higher gate fidelities than the RWA-BS and the RWA protocols.
Also, the RWA-BS protocol has higher fidelities than the RWA one.
{Moreover, for $\gamma=\gamma_{\phi}=2\pi\times0.025$~MHz (which has been realized using superconducting qubits \cite{XiangRMP2013,Gu2017,HuangSCIS2020,Kjaergaard2020,CaiFR2021,WuNjp2021}), the CHRW protocol can reach the threshold of $10^{-4}$ required for quantum error correction \cite{Shor1995Pra,gottesman2010introduction}.
This is difficult for the RWA protocol to match because reaching a higher fidelity requires a longer gate time (see Fig.~\ref{fig1d}), which, however, increases the influence of decoherence.
This is one reason why it is still difficult to experimentally realize a single-qubit geometric gate with fidelity $>99.9\%$ based on the RWA protocols \cite{XuPRL2020,QiuAPL2021,XuOpt2021} (the coherence times of some superconducting qubits are now reaching $1$ ms, as shown in Table \ref{tab4}).}

{
	\renewcommand\arraystretch{1.4}
	\begin{table*}
		\centering
		\caption{Fidelities of geometric quantum gates using superconducting qubits. Coherence properties: energy relaxation time ($T_{1}=1/\gamma$) and
			dephasing time $T_{2}^{*}=1/\gamma_{\phi}$.
			In our protocol, 
			we calculate the gate fidelity by averaging over 10,000 input states, which are uniformly distributed over the Bloch sphere.}
		\label{tab4}
		\begin{tabular}{p{1.9cm}<{\centering}p{5cm}<{\centering}p{1.5cm}<{\centering}p{1.3cm}<{\centering}p{1.3cm}<{\centering}p{1.3cm}<{\centering}p{1.3cm}<{\centering}p{1.5cm}<{\centering}p{1.5cm}<{\centering}}
			\hline
			\hline
			Year \& Ref. & Gate type & $\omega_{q}/2\pi$ (GHz) & $\gamma/2\pi$ (kHz) &  $\gamma_{\phi}/2\pi$ (kHz) & $T_{1}$ \ \ \ \ ($\mu$s) & $T_{2}^{*}$ \ \ \ ($\mu$s) &  Gate time (ns) & Fidelity (\%) \\
			\hline
			\multirow{2}*{2020 \cite{XuPRL2020}} & Single-qubit rotation gates & \multirow{2}*{$\sim$4.61} &\multirow{2}*{$\sim$10} &   \multirow{2}*{$\sim$16} &  \multirow{2}*{$\sim$16} &  \multirow{2}*{$\sim$10} & 80 & $\sim$99.77 \\
			~ & Two-qubit rotation gates &~ & ~ & ~& ~ & ~& 112.8 & $\sim$97.70\\
			\hline
			2021 \cite{QiuAPL2021} & Single-qubit rotation gates & $\sim$5.62 & $\sim$15 & $\sim$29 & $\sim$10.5 & $\sim$5.5 & 100 & $\sim$99.50\\
			\hline 
			2021 \cite{XuOpt2021} & Controlled-NOT gate & $\sim$5.58 & $\sim$12 & $\sim$13 & $\sim$2.1 & $\sim$73   & 205 & $\sim$90.50\\
			\hline
			{Our protocol} & Single- and two-qubit gates & {$\sim$5} &{$\sim$25} &  {$\sim$25}   &  {$\sim$6.4}  &  {$\sim$6.4} & 0.5 & $\gtrsim$99.99 \\
			\hline
			\hline
		\end{tabular}
	\end{table*}
}


\section{Two-qubit gates}
Two-qubit gates can be implemented with the evolution operator
\begin{align}
  \tilde{U}_{\rm{eff}}(t)=\frac{1}{2}\left(\mathbbm{1}^{a}+\sigma_{z}^{a}\right)\otimes\mathbbm{1}+\frac{1}{2}\left(\mathbbm{1}^{a}-\sigma_{z}^{a}\right)\otimes U_{\rm{eff}}(t),
\end{align}
where $\mathbbm{1}$ is the unit operator and the superscript $a$ denotes the additional qubit.
The parameters used for the single-qubit gates can be directly applied to the two-qubit gates.
Hence, when $U_{\rm{eff}}(T)=i\sigma_x$, 
$\tilde{U}_{\rm{eff}}(T)$ corresponds to a CNOT-like gate.
Based on the evolution operator $\tilde{U}_{\rm{eff}}(t)$, we can reversely deduce the corresponding effective Hamiltonian as 
\begin{align}
  \tilde{H}_{\rm{eff}}(t)=i\dot{\tilde{U}}_{\rm{eff}}(t)\tilde{U}_{\rm{eff}}^{\dag}(t)=\frac{1}{2}\left(\mathbbm{1}^{a}-\sigma_{z}^{a}\right)\otimes H_{\rm{eff}}(t).
\end{align}
This effective Hamiltonian is an approximation of the reference
Hamiltonian 
\begin{align}
	\tilde{H}(t)=\frac{1}{2}\left(\mathbbm{1}^{a}-\sigma_{z}^{a}\right)\otimes H(t),
\end{align}
which includes a dipole-dipole interaction $\sigma_{z}^{a}\otimes\sigma_{z}$ \cite{GrajcarPRL2006,NishanenPRB2006,HarrisPRB2009} and a tunable longitudinal coupling
$\sigma_{z}^{a}\otimes\sigma_{x}$ \cite{RicherPRB2016,RicherPRB2017,GarzianoPRL2016,WangPRA2017,WangPRA2018,StassiPRA2018}.
For the CNOT-like gate, the gate time to achieve a fidelity $\geq 99.99\%$ is similar
to that of the NOT gate, i.e., $T\sim 5\pi/\omega$ for the CHRW protocol (see Table \ref{tab2}).
In the presence of decoherence, we assume that the two qubits have the same dissipation rates
and show the gate fidelities in Fig.~\ref{figdiss}(b). As shown, the CHRW and the  
RWA-BS protocols have higher fidelities than those for the RWA protocol, indicating
that employing CR effects can enhance the gate fidelities.

\section{Discussions}
The model discussed here is
generic, so that the proposed proposal can be realized in a wide range of
physical systems.
One of the most promising devices to realize the CHRW protocol can be superconducting circuits
\cite{ClarkeNat2008,YouNat2011,NationRMP2012,XiangRMP2013,Gu2017,WendinRPP2017,KockumNr2019,FornRMP2019,Kjaergaard2020}, 
which have achieved strong interactions \cite{DengPRL2015,Gu2017,KockumNr2019,FornRMP2019}. 
The needed time-dependent detuning or, equivalently,
the time-dependent qubit transition frequency $\omega_{q}$ can be controlled in general, e.g., by Stark shifts. 
The CHRW protocol has a higher speed and a higher fidelity than those of the RWA-BS protocol, 
because the effective Hamiltonian includes high-order terms in the BS shift 
obtained by the Floquet theory \cite{ShirleyPR1965,TuorilaPRL2010,YanPRA2015}.
As a result, even in the weak-driving regime, i.e., $\Omega_{n}(t)\ll \omega$,
the CHRW protocol can achieve a higher fidelity than that for the RWA and RWA-BS protocols, as shown in 
Fig.~\ref{fig1d}. Note that
to avoid the possible excitations to higher-energy levels, 
a system with strong anharmonicity should be used.
{Figure~\ref{fig3} shows the gate infidelities versus the frequency of the second-excited level of the atom. We can see that
to achieve ultrafast quantum geometric gates using our CHRW protocol, a strong anharmonicity
\begin{align}
	\omega_{2}-\omega_{q}\gtrsim 10 \omega,
\end{align}
is needed, where $\omega_{2}$ is the frequency of the second-excited level of the atom.
In general, such a strong anharmonicity is possible using fluxonium and charge qubits \cite{MooijSci1999,DengPRL2015,WendinRPP2017,Gu2017,KockumNr2019,FornRMP2019,NguyenPRX2019,NiemczykNP2010,Forn-DiazPRL2010}.}

\section{Possible implementation with superconducting circuits}\label{AppB}

\begin{figure}
	\centering
	\scalebox{0.7}{\includegraphics{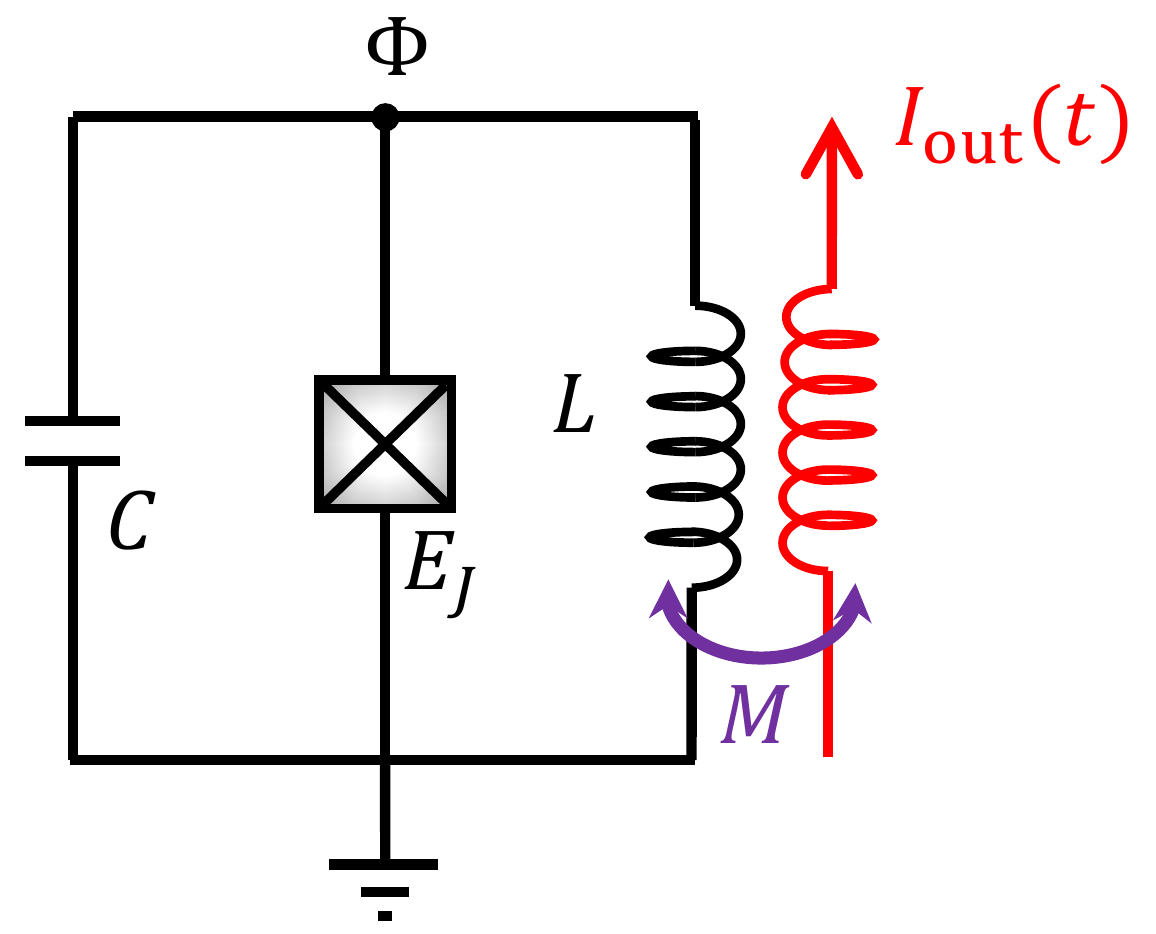}}
	\caption{Circuit of a fluxonium qubit, which is formed by a capacitor $C$, an inductor $L$, and a Josephson junction with Josephson energy $E_{J}$. The qubit is controlled by an external
		current $I_{\rm out}(t)$, which couples to the dimensionless flux variable
		$\Phi$ through a mutual inductance $M$.
	}
	\label{figB1}
\end{figure}

We now present a possible implementation with superconducting circuits for single-qubit gates.
We consider a qubit circuit (see Fig.~\ref{figB1}) formed
by a capacitor $C$, an inductor $L$, and a Josephson junction with
Josephson energy $E_{J}$. The qubit is controlled by an external
current $I_{\rm out}(t)$, which couples to the dimensionless flux variable
$\Phi$ through a mutual inductance $M$. The system Hamiltonian
is 
\begin{align}\label{eqB1}
	H_{\rm{SC}}=&H_{0}+H_{D}, \cr
	H_{0}=&4E_{C}Q^{2}+\frac{E_{L}}{2}\Phi^{2}-E_{J}\cos\left(\Phi+\frac{\Phi_{e}}{\Phi_{0}}\right),\cr
	H_{D}=&\frac{M I_{\rm out}(t)\Phi_{0}}{L}\Phi,
\end{align}
where $Q$ is the dimensionless charge operator, $\Phi$ is the dimensionless flux operator, $\Phi_{0}=\hbar/(2e)$ is the reduced flux quantum, and $M$ is the mutual inductance. Other parameters are $E_{C}=e^{2}/(2C)$ and $E_{L}=\Phi_{0}^{2}/L$.
Here, $Q$ and $\Phi$ satisfy the commutation relations $[\Phi,Q]=i$.
When $\Phi_{e}/\Phi_{0}=\pi$ and $E_{L}<E_{J}$, 
the anharmonicity of the Hamiltonian $H_{0}$ is positive and can be adjusted to a large value \cite{NguyenPRX2019,ZhuPrapp2021}. 

According to the experiment in Ref. \cite{NguyenPRX2019}, we can choose 
$E_{J}/(2\pi)=5~{\rm GHz}$,  $E_{C}/(2\pi)=0.8~{\rm GHz}$, and  $E_{L}/(2\pi)=1.1~{\rm GHz}$, 
resulting in $\omega_{01}/(2\pi)=0.3$~GHz and $\omega_{12}/(2\pi)=3.5$~GHz, where $\omega_{01}$ ($\omega_{12}$) is
the level transition frequency of the first (second) and second (third) eigenstates of $H_{0}$.
The spontaneous emission and dephasing rates are $\gamma=2\pi\times$0.6~kHz and $\gamma_{\phi}=2\pi\times$1.1~kHz \cite{NguyenPRX2019}, respectively.
Then, by accordingly choosing the parameters for the driving Hamiltonian $H_{D}(t)$,
we can obtain the total Hamiltonian in Eq.~(\ref{eq1}), and thus, to realize our protocol.

\begin{figure}
	\centering
	\scalebox{0.43}{\includegraphics{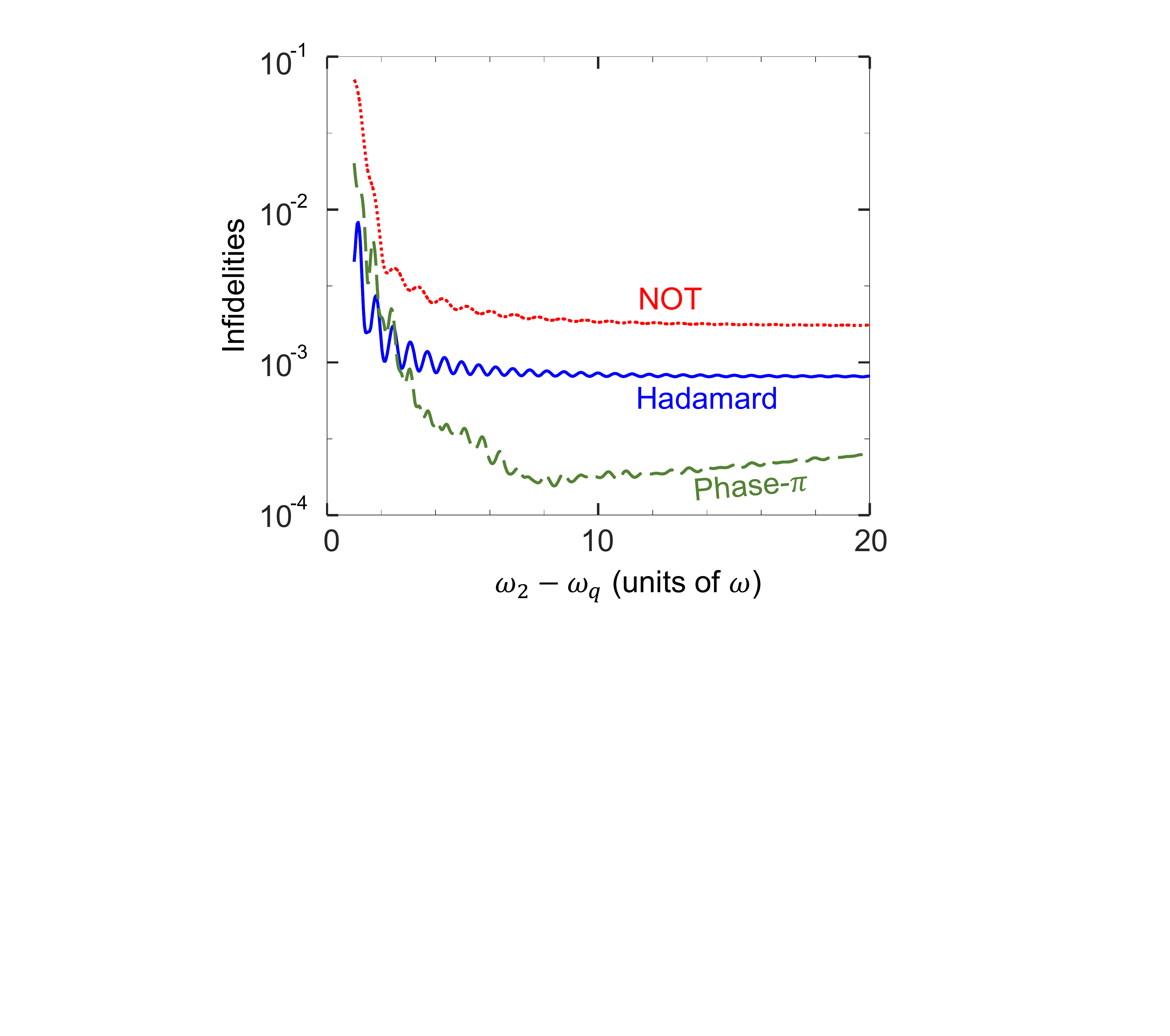}}
	\caption{Gate infidelities ($1-\bar{F}$) of the CHRW protocol when the second-excited level (of frequency $\omega_{2}$) of the atom is considered. Parameters for the plot are listed in Tables~\ref{tab1} and \ref{tab2}.
	}
	\label{fig3}
\end{figure}

\section{Conclusions and outlook}
We show that employing
CR effects (using the CHRW and the RWA-BS protocols) can effectively 
improve the speed and fidelity of geometric quantum computation. 
When CR effects are considered in implementing single- and two-qubit gates,
it is allowed to apply
strong driving fields with amplitudes, which are 
comparable to the atomic transition frequencies.
This significantly improves the gate speed, and, thus, reducing the influence of decoherence.
Moreover, because the CR effects (e.g., the BS shift) are not neglected, 
we can avoid the additional dynamical noise induced by such an effect and further improve the gate fidelity.
No specific design for the driving fields is required in the proposed protocol, and, thus,
the protocol is compatible with most optimal control experimental methods used in previous works.
The proposed protocol also can be generalized to multi-qubit systems using strong couplings. 
Another application of the proposed protocol can be to multi-level (qudit) systems, such as a $\Lambda$-type system \cite{LiuPRL2019,SetiawanPRXQ2021,ZhaoPRA2017,ZhangPRA2019},
which are compatible with other control methods to improve the speed and fidelity of quantum computation.
{All these advantages make our protocol possible to accelerate the previous RWA-based geometric/holonomic quantum computation, so as to improve 
	the fidelity of the computation.}

{
\begin{appendix}
  \section{Invariant-based engineering}	\label{A1}
  For an arbitrary Hamiltonian $H(t)$, we can find a dynamical invariant $I(t)$ satisfying
  \begin{align}\label{R10}
  	i\frac{\partial I(t)}{\partial t}-[H(t),I(t)]=0,
  \end{align} 
  so that the expectation values of $I(t)$ remain constant.
  According to the Lewis-Risenfeld theory \cite{LewisJMP1969}, the solution
  of the Schr\"{o}dinger equation
  \begin{align}
  	i\frac{\partial }{\partial t}|\psi(t)\rangle=H(t)|\psi(t)\rangle,
  \end{align}
  can be expressed as a superposition of the eigenstates of the invariant $I(t)$ as:
  \begin{align}
  	|\psi(t)\rangle=\sum_{n} {c_{n}\exp\left[{i\mathcal{R}_{n}(t)}\right]|\mathcal{I}_{n}(t)\rangle},
  \end{align}
  where $c_{n}$ are time-independent amplitudes, $|\mathcal{I}_{n}\rangle$ are the eigenstates of $I(t)$, and $\mathcal{R}_{n}(t)$ are
  the Lewis-Risenfeld phases:
  \begin{align}
  	\mathcal{R}_{n}(t)=\int_{0}^{t}\left\langle\mathcal{I}_{n}(t')\left|\frac{\partial}{\partial t'}-H(t')\right|\mathcal{I}_{n}(t')\right\rangle dt'.
  \end{align}
  
  For the effective two-level Hamiltonians in Eqs.~(\ref{eq2}) and (\ref{eq8}), the invariant $I(t)$ is found to be
  \begin{align}\label{R15}
  	I(t)=\Xi_{0}\left[
  	\begin{array}{cc}
  		-\cos\beta & i\sin\beta\exp(-i\alpha) \cr
  		-i\sin\beta\exp(i\alpha) & \cos\beta
  	\end{array}
  	\right],
  \end{align}
where $\Xi_{0}$ is a constant to keep $I(t)$ with dimensions of frequency.
 The eigenstates of $I(t)$ are
  \begin{align}
  	&|\mathcal{I}_{+}(t)\rangle=\left[
  	\begin{array}{c}
  		i\sin\left(\frac{\beta}{2}\right)\exp(-i\alpha)\cr
  		\cos\left(\frac{\beta}{2}\right)
  	\end{array}
  	\right],
  	\cr\cr
  	&|\mathcal{I}_{-}(t)\rangle=\left[
  	\begin{array}{c}
  		\cos\left(\frac{\beta}{2}\right)\cr
  		i\sin\left(\frac{\beta}{2}\right)\exp(i\alpha)
  	\end{array}
  	\right].                
  \end{align}
  Thus, we obtain the evolution paths in Eqs. (\ref{eq11}) and (\ref{eq12}), which are 
  \begin{align}
  	|\phi_{\pm}(t)\rangle=\exp[i\mathcal{R}_{\pm}(t)]|\mathcal{I}_{\pm}(t)\rangle.
  \end{align}
  The relationships of the parameters can be obtained by substituting Eqs. (\ref{eq2}), (\ref{eq8}), and (\ref{R15}) into Eq. (\ref{R10}). Thus, we obtain Eqs.~(\ref{eq11})--(\ref{eq6}).

\end{appendix}

}
\begin{acknowledgements}
	We acknowledge helpful discussions with Yi-Hao Kang and Jiang Zhang.
	Y.-H.C. is supported by the Japan Society for the Promotion of Science (JSPS) KAKENHI Grant No.~JP19F19028.
	A.M. is supported by the Polish National Science Centre (NCN)
	under the Maestro Grant No.~DEC-2019/34/A/ST2/00081.
	X. C. is supported by EU FET Open Grant  EPIQUS (899368),  QUANTEK Project (KK-2021/00070),  the Basque Government through Grant No. IT1470-22, the Project Grant 
	PID2021-126273NB-I00 funded by MCIN/AEI/10.13039/501100011033 and by
	``ERDF A way of making Europe'' and ``ERDF Invest in
	your Future''  and the Ram\'on y Cajal program (RYC-2017-22482).
	Y.X. is supported
	by the National Natural Science Foundation of China
	under Grant No. 11575045, the Natural Science Funds
	for Distinguished Young Scholar of Fujian Province under
	Grant 2020J06011 and Project from Fuzhou University
	under Grant JG202001-2.
	F.N. is supported in part by: 
	Nippon Telegraph and Telephone Corporation (NTT) Research, 
	the Japan Science and Technology Agency (JST) [via 
	the Quantum Leap Flagship Program (Q-LEAP), and 
	the Moonshot R\&D Grant No. JPMJMS2061], 
	the Japan Society for the Promotion of Science (JSPS) 
	[via the Grants-in-Aid for Scientific Research (KAKENHI) Grant No. JP20H00134],
	the Army Research Office (ARO) (Grant No. W911NF-18-1-0358),
	the Asian Office of Aerospace Research and Development (AOARD) (via Grant No. FA2386-20-1-4069), and 
	the Foundational Questions Institute Fund (FQXi) via Grant No. FQXi-IAF19-06.
\end{acknowledgements}

\bibliography{references}

\end{document}